\documentclass[preprint2]{aastex}

\newcommand{\footnt}[1]{\footnote{$^)$~#1}$^{)}$}

\slugcomment{~}

\shorttitle{AKARI-CAS}
\shortauthors{Yamauchi et al.}

\begin{document}

\title{AKARI-CAS --- Online Service for AKARI All-Sky Catalogues}

\author{C. Yamauchi, S. Fujishima, N. Ikeda, 
        K. Inada, M. Katano, H. Kataza, S. Makiuti, K. Matsuzaki,
        S. Takita, Y. Yamamoto
        and I. Yamamura}
\affil{
 Institute of Space and Astronautical Science (ISAS),
 Japan Aerospace Exploration Agency (JAXA), 
 3-1-1 Yoshinodai, Chuo, Sagamihara,
 Kanagawa 252-5210, Japan
}
\email{cyamauch@ir.isas.jaxa.jp}

\and

\author{D. Ishihara and S. Oyabu}
\affil{Division of Particle and Astrophysical Sciences, Nagoya University,
Furo-cho, Chikusa-ku, Nagoya 464-8602, Japan}

\label{firstpage}

\begin{abstract}
The {\it AKARI} All-Sky Catalogues are an important infrared
astronomical database for next-generation astronomy that take over the
{\it IRAS} catalog.
We have developed an online service, 
AKARI Catalogue Archive Server (AKARI-CAS), for astronomers.
The service includes useful and attractive search tools and  visual
tools.

One of the new features of AKARI-CAS is cached SIMBAD/NED entries,
which can match AKARI catalogs with other catalogs stored in SIMBAD
or NED.
To allow advanced queries to the databases, direct input of SQL is also
supported.
In those queries,
fast dynamic cross-identification between registered catalogs
is a remarkable feature.
In addition, multiwavelength quick-look images are displayed in the
visualization tools, which will increase the value of the service.

In the construction of our service, we considered 
a wide variety of astronomers' requirements.
As a result of our discussion, we concluded that supporting users' SQL
submissions is the best solution for the requirements.
Therefore, we implemented an RDBMS layer so that it covered important
facilities including the whole processing of tables.
We found that PostgreSQL is the best open-source RDBMS products for such
purpose, and we wrote codes for both simple and advanced searches into the
SQL stored functions.
To implement such stored functions for fast radial search and 
cross-identification with minimum cost, we applied a simple technique 
that is not
based on dividing celestial sphere such as HTM or HEALPix.
In contrast, the Web application layer became compact, and was written in
simple procedural PHP codes.
In total, our system realizes cost-effective maintenance and
enhancements.
\end{abstract}


\section{Introduction}

With the progress of technology, recent astronomical surveys produce
huge object catalogs:
e.g., Sloan Digital Sky Survey \citep[SDSS;][]{yor00},
Two Micron All Sky Survey \citep[2MASS;][]{skr06}, 
UKIRT Infrared Deep Sky Survey \citep[UKIDSS;][]{law07}, etc.
It is not realistic for many astronomers to handle and analyze 
a number of catalogs in their local computers.
Therefore, services to access the data via network have to be developed
and could be used for a wide variety of astronomers' requirements.
In addition,
it is desired that the services have application program
interfaces (APIs) to develop services, 
since it is not efficient that all service providers have all
observational data.

In modern Web-based database solutions for object catalogs --- for example,
NASA/IPAC Infrared Science Archive 
\citep[IRSA;][]{ber00},
VizieR \citep{och00},
and Virtual Observatory \citep[VO;][]{sza01} --- portal sites
are famous services.
One of their common features is that we can use rich 
graphical user interfaces (GUIs) and
a high-level programming interface based on VO standards
for our various demands.
On the other hand, 
one of the most advanced services, 
SkyServer of SDSS \citep[][]{tha04},
established a new basic model of a modern Web-based database solution
for a huge catalog.
SkyServer provides not only basic search tools such as radial search
but also advanced search tools such as the SQL search, CAS jobs, etc.
Compared with the former services,
SkyServer of SDSS is developed 
with focus on the 
availability of flexible SQL-based programming interface,
rather than supporting rich GUIs and high-level interfaces.
Such SQL-based programming interfaces will become very important for
online services dedicated to long-life survey database,
since astronomers often need
advanced search or their own programs to analyze catalogs and
observational data for various unique science themes.

Our archive server AKARI-CAS%
\footnt{AKARI-CAS has been developed based on open source softwares.
If readers are interested in our system,
we can show our source codes.
Please contact us about them.}
has been developed to provide
Web-based access tools for AKARI All-Sky Catalogues.
The tools have been developed following the concept of SDSS SkyServer,
and users can search, match up, and browse stored data using the
provided tools.
In our service, there are tools based on our own ideas and
implementation techniques.
Reporting them in this article will be worthwhile information 
that is applicable to developments of future astronomical Web-based
database systems.

In this article, we mainly discuss the implementation and design of
a Web-based service for astronomical catalogs.
In addition, our article will be helpful to develop basic search
tools with minimum cost and to set up RDBMS security
when the system allows users direct input of SQL statements.
This article is organized as follows:
In \S \ref{overview_catalogs},
we briefly introduce AKARI All-Sky Catalogues.
In \S \ref{overview_akari-cas},
we describe an overview of our catalog archive service.
In \S \ref{overview_technical},
we discuss the basic direction of our technical design.
Technical reports for RDBMS and Web application layers
are given in
\S \ref{implementation_rdbms} and
\S \ref{implementation_webapp}, respectively.
A summary is given in \S \ref{conclusions}.

\section{AKARI All-Sky Catalogues Overview}
\label{overview_catalogs}

AKARI \citep{mur07}
is the second Japanese space mission for infrared astronomy,
which has been developed by the 
Institute of Space and Astronautical Science (ISAS) at 
Japan Aerospace Exploration Agency (JAXA) 
and collaborators.
The AKARI mission is an ambitious plan to make an all-sky survey with
better sensitivity, higher spatial resolution, and wider wavelength
coverage than those of {\it IRAS} \citep[][]{hel88}.
AKARI is equipped with 
a telescope of 68.5 cm effective aperture and two instruments:
the Far-Infrared Surveyor \citep[FIS;][]{kaw07}
for the far-infrared observations and
the Infrared Camera \citep[IRC;][]{ona07}
for the near and mid-infrared wavelengths.
AKARI was launched by the M-V8 vehicle on 2006 February 22.
AKARI All-Sky Survey started in 2006 May and was completed in 2007 August.

AKARI All-Sky Catalogues are produced based on the obtained all-sky
survey data.
The public release of the first version of catalogs was made 
on 2010 March 30.
This release consists of two infrared catalogs: 
the FIS Bright Source Catalogue \citep[BSC;][]{yam10}
with 427,071 objects observed in the four far-infrared wavelengths, and
the IRC Point Source Catalogue \citep[PSC;][]{ish10}
including 870,973 objects in the two mid-infrared wavelengths.
The catalog set covers more than 98\% of the sky.
The information in the catalogs contains the object identifier, J2000 position,
flux data, quality flags, and observational information.
The catalog files can be obtained as a FITS file or a text formatted file,
and the release notes are also provided.%
\footnt{See http://www.ir.isas.jaxa.jp/AKARI/Observation/PSC/\\Public/
for details.} 
There is no information for cross-identification between FIS BSC
objects and IRC PSC objects in the released catalogs.
Note that image data of the AKARI All-Sky Survey are not yet
publicly released.


\section{AKARI Catalogue Archive Server}
\label{overview_akari-cas}

\begin{figure*}[!t]
\epsscale{1.7}
\plotone{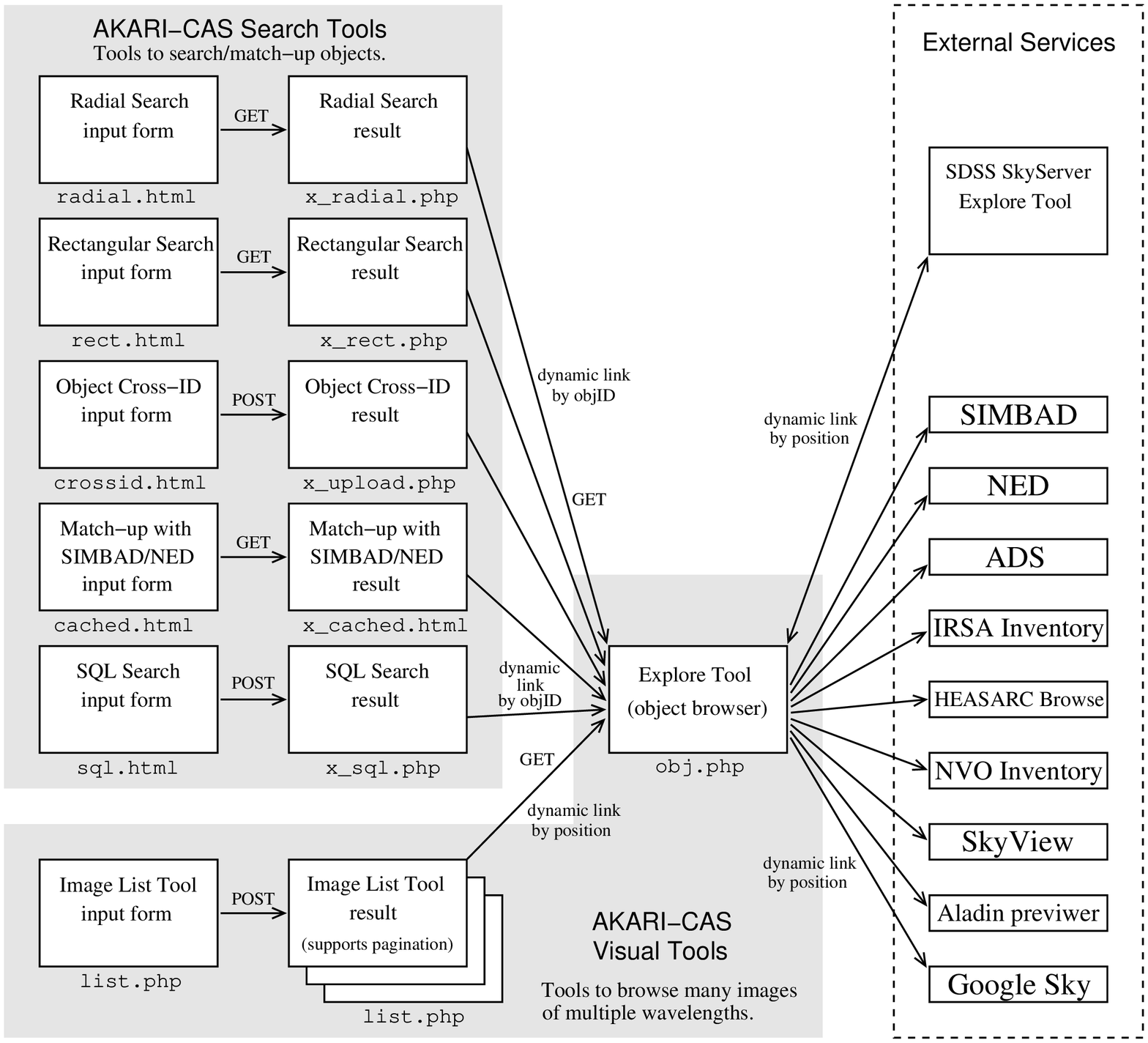}
\caption{Overview of page transitions of AKARI-CAS and
 other Web services.
 Links to Explore Tool are dynamically created
 in the results of Search Tools and Visual Tools.
 Explore Tool dynamically creates links to an external Web site by
 right ascension and Declination.
 Thus, users can visit various external sites for
 all searched objects and images via Explore Tool.
 The SDSS project team kindly updated their Explore Tool of SkyServer
 to append links to AKARI-CAS Explore Tool.
}
\label{fig:page_transition}
\end{figure*}

\begin{figure*}[!t]
\epsscale{1.7}
\plotone{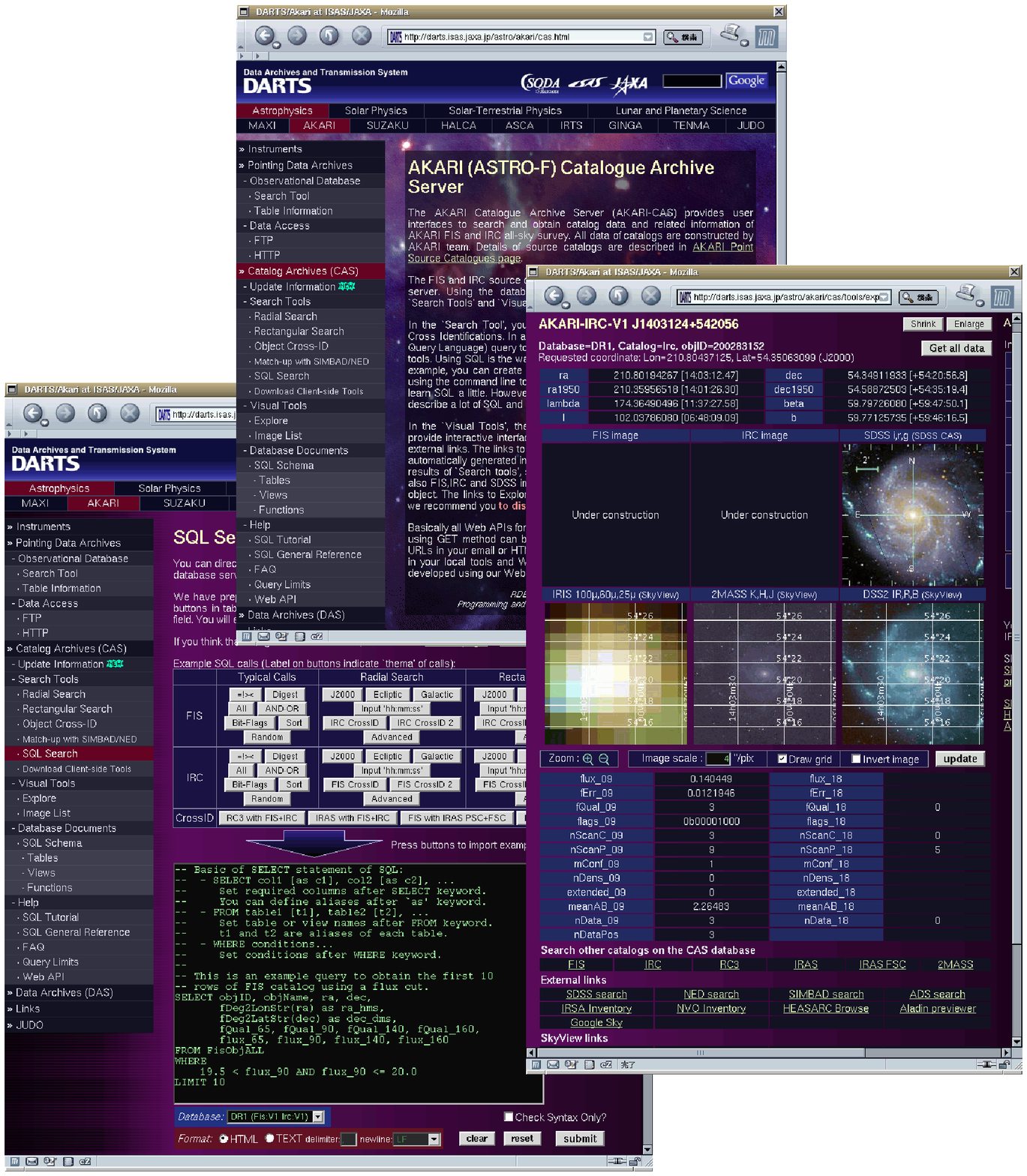}
\caption{Top page (top), SQL Search (left), and Explore Tool (right)
 of AKARI-CAS.
The links to all tools and documentations are always 
displayed on the left side of all pages, except Explore Tool.
Pages are categorized into Search Tools, Visual Tools,
Database Documents, and Help.
Displaying quick-look images of multiwavelengths and
supporting direct input of an SQL statement are
distinctive features of our service.
Our service covers various demands from general users to
power users.
}
\label{fig:top_page}
\end{figure*}


AKARI Catalogue Archive Server (AKARI-CAS) provides basic tools for
various astronomical studies with the AKARI catalogs.
For the ground design of our service, we referred to the SDSS SkyServer
and combined our new idea into it.
Figure \ref{fig:page_transition} shows the relation of AKARI-CAS Web
tools and external services.
The tool at the center of this figure is the object browser, Explore Tool,
that plays the leading role in our Web interface.
Basically, hyperlinks to Explore Tool are created
in {\em all} search results of AKARI-CAS, and
Explore Tool dynamically prepares hyperlinks to external sites, i.e.,
SIMBAD%
\footnt{See http://simbad.u-strasbg.fr/simbad/.}
\citep[][]{wen00}, NED%
\footnt{See http://nedwww.ipac.caltech.edu/.}
\citep[][]{hel91}, etc.
It is possible to create a permalink to the Explore Tool by an object
position.
The links generated by external services
enable users to refer detail of an AKARI object.
This design is essential to attain an interoperability feature of
our Web interface.

Figure \ref{fig:top_page} shows screenshots of the top page, 
Explore Tool page, and  SQL Search page, 
which supports direct SQL input.
It is important to note that the APIs of AKARI-CAS are open to the public.
Both astronomer's scripts and Web service programs can call the APIs.
Tools and documentation pages are arranged 
on the menu placed on the left side of Web pages. 
As shown in Figure \ref{fig:top_page},
our Web tools are classified into two categories:
Search Tools and Visual Tools.
We briefly explain them
in \S\ref{overview_search_tools} and \S\ref{overview_visual_tools}
after discussing stored data in \S\ref{overview_stored_data}.
Our Web APIs are summarized in
\S\ref{overview_webapis}.

\subsection{Stored Data}
\label{overview_stored_data}

We registered the AKARI catalogs and several external data that are
expected to be important for astronomical studies into our databases.
The data used through our Web tools are as follows:
\renewcommand{\labelenumi}{\arabic{enumi}.}
\begin{enumerate}
 \item
      AKARI All-Sky Survey FIS Bright Source Catalogue (FIS BSC) version 1.0
 \item
      AKARI All-Sky Survey IRC Point Source Catalogue (IRC PSC) version 1.0
 \item
      Third Reference Catalog of Bright Galaxies \citep[RC3;][]{vau91}
 \item
      {\it IRAS} Point Source Catalog (IRAS PSC) version 2.1
 \item
      {\it IRAS} Faint Source Catalog (IRAS FSC) version 2.0
 \item
      2MASS Point Source Catalog (2MASS PSC)
 \item
      Cached object list of SIMBAD
      matched up with AKARI BSC/PSC objects
 \item
      Cached object list of NED
      matched up with AKARI BSC/PSC objects
\end{enumerate}
The cached SIMBAD/NED object list contains entries found around each AKARI
object within a 1.0$'$ radius.
The update of the cached list in the databases will be done every few months.

\subsection{Search Tools}
\label{overview_search_tools}

\begin{figure*}[!t]
\epsscale{1.7}
\plotone{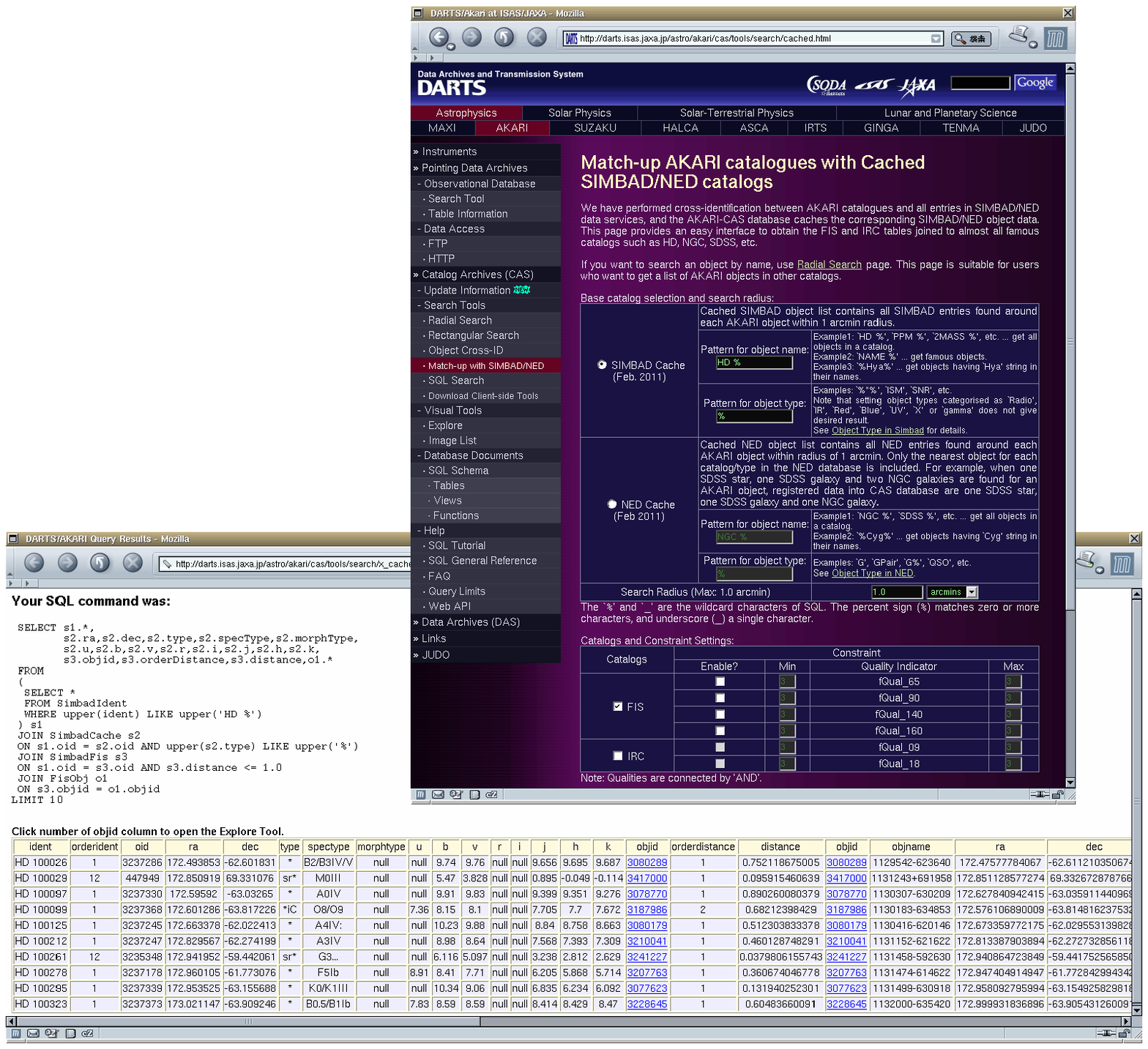}
\caption{Input form and a result of Match up with SIMBAD/NED tool of AKARI-CAS.
With an easy GUI operation, users can obtain
a list of any object catalogs in SIMBAD/NED joined with AKARI catalogs.
This example shows the Henry Draper (HD) Catalogue joined with AKARI FIS BSC.
}
\label{fig:matchup_simbad}
\end{figure*}

Search Tools consists of following tools
to search AKARI catalogs with various criteria.
Positions (longitude, latitude) can be set by J2000, B1950, ecliptic or
Galactic coordinates in all Web-based tools.
\renewcommand{\labelenumi}{\arabic{enumi}.}
\begin{enumerate}
 \item
      The Radial Search
      tool provides an easy interface to search FIS and/or IRC objects
      within a radius of a position, which can
      specified by direct coordinate input or object name resolved by
      SIMBAD or NED.
 \item
      The Rectangular Search 
      tool is used to
      search FIS and/or IRC objects inside a rectangle defined by two
      longitude, latitude pairs.
 \item
      The Cross-Identification
      tool provides an interface to search
      FIS and/or IRC objects with a user defined list of object
      positions. 
      In the result, FIS objects and IRC objects are crossly identified.
 \item
      For the match up with SIMBAD/NED catalogs,
      the results of cross-identification between AKARI objects and
      SIMBAD/NED objects were registered into CAS database.
      Users can easily obtain AKARI catalogs joined to cached catalogs
      obtained from SIMBAD/NED such as HD, MACHO, 2MASS, NGC, SDSS,
      etc. 
      Figure \ref{fig:matchup_simbad} shows the input and result display
      pages of this tool;
      the Henry Draper (HD) Catalogue is matched with AKARI FIS-BSC.
 \item
      With the SQL Search tool,
      users can directly submit an SQL query to RDBMS of the AKARI-CAS
      using this tool.
      The SIMBAD/NED cached data,
      RC3 catalog, {\it IRAS} catalogs and 2MASS PSC can be used in this tool.
      Users can perform fast cross-identification between AKARI
      catalogs and these catalogs.
 \item
      Access to the SQL Search is also possible from 
      users' command-line programs.
      The Download Client-side Tools
      page provides the information for client-side utilities 
      written in Python or Interactive Data Language (IDL)
      to perform an SQL search on the command line.
      They can be included in users' programs for advanced analysis.
      Any user can download these client-side utilities from
      this page.

\end{enumerate}

\subsection{Visual Tools}
\label{overview_visual_tools}

Tools with quick-look images are categorized in Visual Tools.
Supported quick-look images are currently
SDSS, IRIS(Improved Reprocessing of the {\it IRAS} survey),%
\footnt{See http://www.cita.utoronto.ca/$^\sim$mamd/IRIS/.} 
2MASS, and DSS2.
\renewcommand{\labelenumi}{\arabic{enumi}.}
\begin{enumerate}
 \item
      The Explore Tool
      page provides a useful interface to access 
      catalog data of AKARI, RC3, {\it IRAS} and 2MASS,
      and quick-look images.
      External links such as ADS, IRSA, SkyView,%
\footnt{http://skyview.gsfc.nasa.gov/} 
      etc.
      are automatically generated on this page.
      Users can specify an object by its position,
      object name, or object identification number (ObjID).

 \item
      After uploading the position list, users can view 40 quick-look
      images per page with
      Image List Tool.
      Links to the Explore Tool are also generated.
      This tool does not access any catalogs in our RDBMS.
\end{enumerate}

\subsection{Web APIs}
\label{overview_webapis}

Search Tools and Visual Tools have open Web APIs (GET or POST).
Such APIs are important for not only astronomical
study but also for software development.
For example, the Herschel Observation Planning Tool (HSpot) accesses
our API of Radial Search 
via network and plots AKARI objects in the HSpot GUI.
Explore Tool of SDSS SkyServer has two dynamic links to
our Explore Tool after the update of 2010 April by the SDSS project team.
The APIs with the GET method can be used as permalink.
A URL to each query can be given as text in e-mails or in HTML documents,
which is useful for the users to communicate with the collaborators.

Web programs in Search Tools return result tables in HTML or text
(CSV) format.
A VO interface (e.g., VOTable, TAP, etc.) will be supported in the
future.
See Appendix A
for more information on the APIs.

\section{Overview of Technical Design of CAS}
\label{overview_technical}

Our service shall be friendly for many levels of users, from beginners to
expert users.
In this section,
we describe our direction of technical design for such requirements.

\subsection{Roles of RDBMS Layer and Web Application Layer}
\label{roles_layers}

When we develop a Web-based database system, we usually design two layers;
RDBMS and Web application.
The definition of the role of each layer is essential for our development,
since the flexibility of service is almost determined by it.

We expect that beginners want easy-to-use GUI tools, while expert users
demand the direct access to databases.
Moreover, if our tables have complicated relations, the SQL interface
may be indispensable for advanced analysis.
Therefore, 
it is unavoidable that our service accepts users SQL statements like
SDSS SkyServer.
In this case, we should prepare a couple of stored functions 
that execute SQL statements that are indispensable for typical searches 
and documents that press users for use of such functions,
since users do not always have enough skill with RDBMS.
Preparing some stored functions for astronomical
studies such as conversion of coordinates between sexagesimal and
degrees, etc., is helpful for users to submit SQL statements.

After the preceding discussion,
we define the role of each layer as follows:
\renewcommand{\labelenumi}{\arabic{enumi}.}
\begin{enumerate}
 \item
      The RDBMS Layer
      covers the entire processing of tables, and conversion of
      coordinate and physical values.
      Codes for both simple and advanced searches are built-in SQL
      stored functions.
      Using stored function, users can use SQL statements without
      knowledge about indices of tables.
      
 \item
      The role of 
      the Web application layer
      is only to accept and check the
      users' input and to generate simple SQL statements.
      To support the transmission of huge table data,
      this layer should not have any buffer to cache the data.
      To extend possibility of various applications,
      we apply the stateless design to this layer.

\end{enumerate}

We expect that the preceding direction can also minimize the cost
of maintenance of our service.

\subsection{Used Software Products}

We introduce the software products used to construct AKARI-CAS and
explain why we choose them.

In the RDBMS Layer, we employ PostgreSQL-8.4.
	    As we described in \S \ref{roles_layers},
	    our AKARI-CAS allows direct input of SQL statement from
	    users.
	    It means that the RDBMS should have enough security
	    configuration system and high coding flexibility of stored
	    functions in SQL, procedural language and C.
	    It is known that some stored functions of SDSS SkyServer
	    return multiple rows and are very useful to construct
	    SQL statements for users.
	    Therefore, our stored functions of RDBMS should be able to
	    return multiple rows.
	    In addition, to construct the mirror site or
	    to apply our codes to other projects flexibly,
	    it is desirable that we use an open source RDBMS software.
	    After our investigation of some RDBMS softwares,
	    we found that PostgreSQL perfectly satisfies above requirements.
	    
We also use
	    wcstools-3.8%
	    \footnt{http://tdc-www.harvard.edu/wcstools/}
with PostgreSQL.
	    To convert coordinates (e.g., from Galactic to J2000)
	    in SQL statements, we wrote
	    some stored functions in C
	    and register them into our databases.
	    The source codes in the stored functions call
	    some routines of wcstools.

In the      Web application layer,
we use PHP-5.3.
	    As explained in \S \ref{roles_layers}, we expected that our
	    implementation of Web application can be concise, since
	    important facilities were built in the RDBMS layer.
	    Therefore, we had flexibility in our selection of languages
	    that can be used for the Web application and
	    have libraries for accessing RDBMS.
	    Perl CGI has been broadly used since more than 10 years ago,
	    however, Java/Tomcat and PHP become very popular for
	    building Web application in recent years.

	    According to our investigations about developments of 
	    simple stateless Web applications at ISAS, we found two facts:
\begin{enumerate}
 \item
      The efficiency of development in PHP is better than that
      in Java, since we do not have to compile our sources in PHP.
 \item	    
      PHP does not force us to use class, therefore,
      we can rapidly write codes in procedural style.
\end{enumerate}

	    We concluded that PHP is more suitable for our development,
	    and confirmed that PHP applications run fast
	    enough compared with other Web application platforms.

\section{Implementation of RDBMS layer}
\label{implementation_rdbms}

The key features of our implementation are a number of kinds of stored
function, fast radial search, and cross-identification using
indexing 
that is not based on
Hierarchical Triangular Mesh%
\footnt{See http://skyserver.org/HTM/.}
\citep[HTM;][]{kun01}
or
HEALPix%
\footnt{See http://healpix.jpl.nasa.gov/.}
\citep{gor05},
and the security configuration for direct input of users SQL
statements.
We mainly explain these points in this section.

\subsection{Assignment of ObjIDs to Registered Catalogs}

\begin{table}[!b]
\begin{center}
\caption{Assignment of objIDs to catalog objects.
For example, {\tt objID} = {\tt 200123456} is
123,456th object of the IRC catalog of version code 20.
Note that this assignment might be changed.
}\label{tbl:objid}
\end{center}
\begin{center}
\renewcommand{\tabcolsep}{2.4pt}
\begin{tabular}{rcl}
\tableline\tableline
\multicolumn{1}{c}{objID} &
\multicolumn{1}{c}{\small range of {\tt VV}} &
\multicolumn{1}{c}{catalog} \\
\tableline
   {\tt \footnotesize VVxxxxxx} & {\tt 01}-{\tt 49} & FIS catalog(s)
 \\
   {\tt \footnotesize VVxxxxxx} & {\tt 50}-{\tt 99} & {\footnotesize other AKARI-related catalogs}\\
  {\tt \footnotesize VVxxxxxxx} & {\tt 10}-{\tt 39} & IRC catalog
 \\
  {\tt \footnotesize VVxxxxxxx} & {\tt 40}-{\tt 49} & external catalogs (large)\\
  {\tt \footnotesize VVxxxxxxx} & {\tt 50}-{\tt 99} & 2MASS PSC\\
 {\tt \footnotesize 1xxxxxxxxx} & - & Reserved\\
 {\tt \footnotesize 20xxxxxxxx} & - & Reserved\\
 {\tt \footnotesize 21VVxxxxxx} & {\tt 01}-{\tt 46} & external catalogs (small)\\
\tableline
\end{tabular}
\end{center}
\end{table}

We defined a unique number for each object of each catalog before
registering them into databases.
We named the numbers ObjID.
Table \ref{tbl:objid}
shows the current assignment rule of the ObjIDs.

In Figure \ref{fig:page_transition}, we show that Explore Tool is the
main tool in our service.
The dynamic links from the search result pages to Explore Tool are
created using ObjIDs.
Then, Explore Tool parses the ObjID in the arguments of the GET method,
accesses the appropriate table, and displays required data.

Using ObjID in links to Explore Tool,
multiple arguments such as instrument, position, etc. in URL are not
required in such links.

\subsection{Table Relationships}
\label{table_relationships}

\begin{figure*}[!t]
\epsscale{1.7}
\plotone{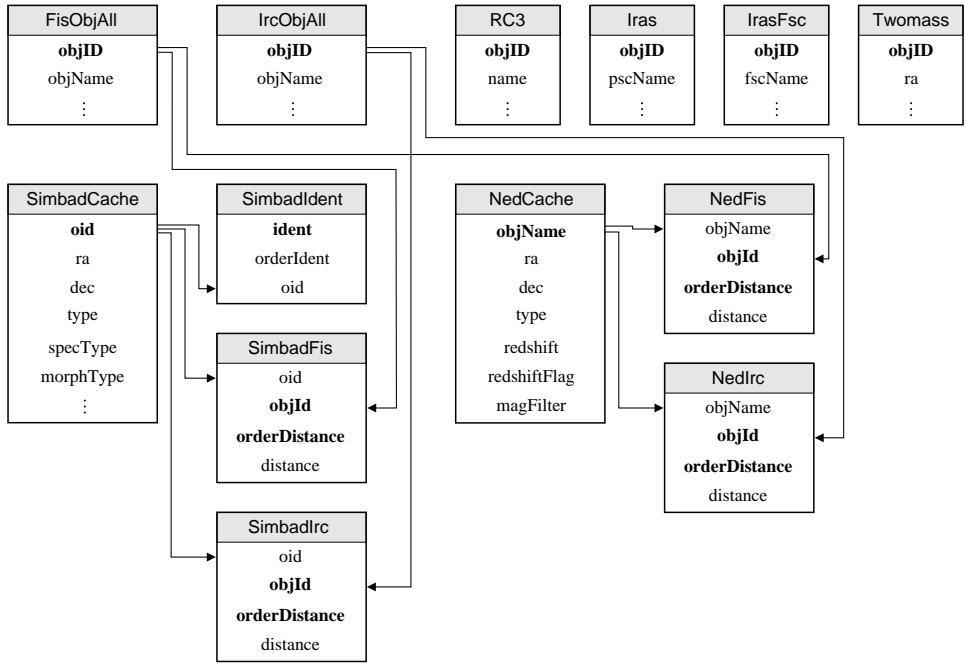}
\caption{Relationships of the tables in AKARI-CAS RDBMS.
The upper five table (object catalogs)
have no information for cross-identification between
catalogs; however,
such cross-identification is dynamically done
using an SQL statement with stored functions.
}
\label{fig:table_relation}
\end{figure*}

As we outlined in \S\ref{overview_akari-cas}, we registered AKARI
catalogs and other important data into our databases to support users'
various demands.
We show the relationships between the tables in Figure
\ref{fig:table_relation}.
The upper five tables shown in this figure are object catalogs, which have
no information for cross-identification between catalogs.
Such cross-identification is dynamically carried out using stored
functions.
Details are described in \S\ref{cross_id}.

On the other hand, SIMBAD/NED tables have static information of
cross-identification.
AKARI catalogs and cached SIMBAD/NED entries are easily joined to 
each other using {\tt JOIN} or {\tt LEFT JOIN} keywords.
We also prepared some views such as SimbadCacheAll, NedCacheAll,
etc.
They are useful to join SIMBAD/NED information
to the searched results of AKARI catalogs.
Details of tables and views are given in our
Database Document pages.

\subsection{Stored Functions}
\label{stored_functions}

We created variety of stored functions for our databases.
They can be categorized as follows:
\renewcommand{\labelenumi}{\arabic{enumi}.}
\begin{enumerate}
 \item
      String processing.
 \item
      Unit conversions of physical values such as
      AB magnitude and flux (in janskys).
 \item
      Conversion of coordinate between sexagesimal and degrees.
 \item
      Simple calculations of positional values such as
      separations between two positions.
 \item
      Low-level functions using wcstools routines.
 \item
      Stored functions that return multiple rows such as
      radial search, rectangular search, etc.
      They have indispensable SQL statements for users
      to run typical searches
      that heavily depend on the indices of the tables.
\end{enumerate}
Together with several built-in stored functions of PostgreSQL, various
processing can be done in an SQL statement in our databases.
Details are shown on our CAS document page.

To implement the preceding stored functions,
we used PL/pgSQL, C, and SQL from provided languages in PostgreSQL.
PL/pgSQL is a procedural language that is suitable to return a value from
numerical calculations and/or SQL executions and is typically used for
function 1--4 in our system.
For example, {\tt SELECT fLonStr2Deg('12:02:00.00')} returns
{\tt 180.5} in degrees.

Stored functions written in C call routines of wcstools, 
as shown in function 5.
In addition, C is powerful enough to obtain better speed for performing
a sequential scan with calculations in RDBMS.
Some functions in function 4 are used with such a sequential scan.
For example, we wrote {\tt fDistanceArcMinXYZ()} in C, which calculates
an angle between two objects.
This stored function is frequently used in the radial search and is
essential for system performance.
According to our test, we found that the radial search routine
with {\tt fDistanceArcMinXYZ()} written in C performed double speed in
several degrees, compared with that in PL/pgSQL.

SQL is used for SQL executions, can return multiple rows,%
\footnt{Before creating a stored function that returns multiple rows,
we have to register a new type.} 
and is used for the function 6.
For example, {\tt SELECT * FROM fGetNearbyObjEq('Fis', 195.5, 2.5, 40.0)}
returns the following result:
\begin{center}
{\tiny
\renewcommand{\tabcolsep}{1.0pt}
\begin{tabular}{ccccc}
{\tt objid} & {\tt cx} & {\tt cy} & {\tt cz} & {\tt distance}\\
{\tt 3151469} & {\tt -0.964260661127699} & {\tt -0.261334534370321} & {\tt 0.0436536200906656} & {\tt 20.1372052103085}\\
{\tt 3204984} & {\tt -0.962002401977625} & {\tt -0.267611949457189} & {\tt 0.0541776992590366} & {\tt 36.4431406886059}\\
\end{tabular}
}
\end{center}
This is a result of searching objects within 40.0$''$ radius
from (right ascension, declination) = (195.5, 2.5).
Generally, the preceding result is used to join FIS object table or view with it.

In total, about 2000 lines are newly written in PL/pgSQL, C, and SQL,
including comments for our stored functions.

\subsection{Radial Search}
\label{radial_search}

Radial search is the most typical query in the astronomical 
Web-based database services.
However, this search is somewhat unusual for RDBMS, and special methods
to create a one-dimensional index (HTM and HEALPix) have been devised.
We did not try such methods, but used a more cost-effective
method.

The object tables in our databases have columns of unit vectors 
({\tt cx}, {\tt cy}, and {\tt cz})
presenting the J2000 source positions.
In order to force the radial search to be fast enough,
we created composite indices on all object tables,
such as\\
{\footnotesize \verb|CREATE INDEX fisobjall_xyz ON FisObjAll (cx,cy,cz)|,}\\
and wrote stored functions that catch objects within a cube using
index-scan and then drop objects outside the strict search circle on the
celestial sphere.%
\footnt{In our tests, we have confirmed that this method is
useful for radial search and cross-identification of huge catalogs such
as 2MASS PSC. We will report the result in another article.
}

We tested the performance on an Opteron2378 (2.4GHz) PC and reached a
result of less than 0.01 s
for a radial search of the IRC catalog within 20$'$ radius from the
Galactic center (177 objects are returned).
More than 0.4 s was consumed without the index; therefore, the
result demonstrated the usefulness of our method.
Note that this test was made in the psql interactive terminal with
the following SQL statement:\\
{\footnotesize \verb|SELECT count(*)|}\\
{\footnotesize \verb|FROM fGetNearbyObjCel('irc', 'gal', 0.0, 0.0, 20.0);|}\\
after the \verb|\timing| command.

\subsection{Cross-Identification between Registered Catalogs}
\label{cross_id}

As shown in \S \ref{table_relationships}, tables of object catalogs do
not have columns of ObjIDs of other catalogs.
We created the {\tt f}{\it name}{\tt GetNearestObjIDEq()} function to get
the ObjID of the nearest object from a given position in specified
object catalog ({\it name}) by performing a radial search.
For example, 
{\tt fIrcGetNearestObjIDEq(180.67, 1.977, 1.0)}
searches the nearest object from the position (180.67, 1.977) within
1.0$'$ in the IRC catalog.
A cross-identification between registered catalogs is carried out by
an SQL statement to repeat calling the stored function with appropriate
arguments.

The algorithm of the stored function is the same as that used in the radial
search (\S \ref{radial_search}) and
has the great advantage of performing the cross-identification, since
the cost of
calculations before executing an index scan for a radial search is quite
small compared with that based on techniques with dividing celestial
sphere such as HTM or HEALPix.

For example, the following SQL statement searches for FIS objects of
fQual\_65 = 3 and joins the FIS search result with the IRC catalog:\\
\verb|SELECT p.*, q.*|\\
\verb|FROM|\\
\verb|(|\\
\verb|  SELECT *|\\
\verb|  FROM FisObj|\\
\verb|  WHERE fQual_65 = 3|\\
\verb|) p|\\
\verb|LEFT JOIN IrcObj q|\\
\verb|ON fIrcGetNearestObjIDEq(p.ra, p.dec, 1.0)|\\
\verb|   = q.objID;|\\
In this case, the radius to match up objects between the FIS and IRC
catalogs is 1.0$'$, which is set in the last argument of
the {\tt f}{\it name}{\tt GetNearestObjIDEq()}
function.
Some examples of SQL statements for cross-identification are
given on the SQL Search page.

In our test using a 2.4GHz Opteron PC, the elapsed time was
27 s to match up all FIS objects with all IRC objects within a 15
$''$ radius.
This means that about
15,800
radial searches are processed per second.
The following is the SQL statement for this search:\\
{\footnotesize \verb|SELECT count(fIrcGetNearestObjIDEq(ra, dec, 0.25))|}\\
{\footnotesize \verb|FROM fisobj;|}

\subsection{Cross-Identification between a Registered Catalog and Users List}
\label{cross_id_users}

To perform a cross-identification between a catalog table and the result of
an SQL statement, we created a stored function using PL/pgSQL, 
{\tt fGetCrossIdResultEq()},
which returns a table of results.

The function {\tt fGetCrossIdResultEq()} takes four arguments:\\
\verb|fGetCrossIdResultEq('fis',|\\
\verb|      'SELECT * FROM upload', 3.0, false)|\\
The first argument is the name of catalog, 
the second one is an arbitrary SQL
statement that returns multiple rows in the form of
{\tt (id\_x INT4, name\_x VARCHAR, ra\_x FLOAT8, dec\_x FLOAT8)},
the third one is the radius to search in arcminutes, and the last one is
a flag to
indicate the search scope: 
{\tt false} means to search the nearest object only, and 
{\tt true} means to search all nearby objects.

In the actual implementation of our cross-identification tool, uploaded users'
position lists are converted into J2000 coordinates, and they are
registered into a temporary table upload.
Then {\tt fGetCrossIdResultEq()} is called.
In this case, the second argument should be {\tt 'SELECT * FROM upload'} to
scan a temporary table upload.
An SQL statement to scan existing tables can also be set to the second argument.%
\footnt{However, our Web interface does not currently support such a use.}

The implementation of {\tt fGetCrossIdResultEq()} is not complicated.
After opening a cursor%
\footnt{In PL/pgSQL, cursor means a pointer to a row.}
for the J2000 position list returned from an SQL execution of the second
argument, each position is scanned and {\tt fGetNearbyObjEq()} is
performed.
The results of {\tt fGetNearbyObjEq()} are pushed into a temporary
buffer, and the function returns the buffer after closing the cursor.
Thus, the code of {\tt fGetCrossIdResultEq()} is a typical PL/pgSQL
application.

\subsection{Security Configuration}
\label{pgsql_security}

\begin{figure*}[!t]
\epsscale{1.7}
\plotone{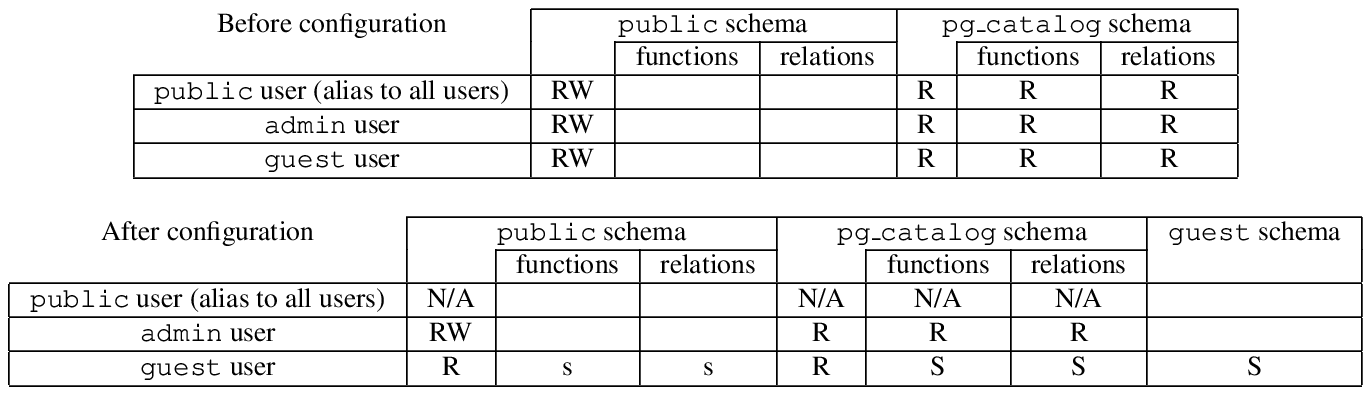}
\caption{Permission of database before and after running
script for security configuration.
The letters R and RW indicate read-only and read-write permissions,
respectively.
The selected grant is set for S functions and relations.
Our script runs under the postgres user and
sets R, RW, and S permissions.
The s permissions are set by owner of functions or relations.
}
\label{fig:security}
\end{figure*}

Built-in stored functions of PostgreSQL are so powerful
that some of them can call 
low-level functions of the operating system.
Therefore, we have to be careful with
the security setup of the database
before allowing the direct input of SQL statements by users.

To configure the security setup, 
we have to understand the concept of a {\it schema} of PostgreSQL.
A schema of PostgreSQL is a section in which
relations (tables and views) and functions exist.
It is like a directory in which files are located in UNIX systems.
Schemata pg\_catalog and public are created in a new database
by default.
The pg\_catalog schema has built-in relations and functions.
Additional relations are registered into the public schema.

PostgreSQL provides two users, postgres and public, by default.
User postgres is the super-user, and public is the alias to all users.
We appended admin user to create relations and functions and
guest user to access the database from Web applications.

Just after creating a database,
all users can create functions and relations in the public schema and
can use or access functions and relations in the pg\_catalog schema.
This is not secure at all for our purpose, and so 
we made a script to set up the right security configuration of our database.
We show the permission of a database 
before and after running the script in Figure \ref{fig:security}.
The permission of the s symbol (i.e., tables and functions for our astronomical data)
in Figure \ref{fig:security} is set by the admin user.
See Appendix B for details of the script.

Users input such as SQL statements should be tested in the Web
application layer.
Even if Web applications have vulnerability, the preceding configuration shall
protect our database from malicious inputs.

\section{Implementation of Web Application layer}
\label{implementation_webapp}

Web application programs in AKARI-CAS are written in simple
procedural PHP codes without object-oriented technique.
Although PEAR::MDB2 and PEAR::HTTP\_Request2 libraries are used in
our codes, we did not use 
any frameworks for PHP.
In total, about 4300 lines are newly written in PHP including comments
and HTML.

We mainly explain some ideas to construct our Web application
effectively in this section.

\subsection{Search across Multiple Catalogs}
\label{search_multiple_catalogs}

Considering the future extension of services in ISAS/JAXA and other
institutes, we implemented
Radial Search and Rectangular Search of our service 
so that they can search the FIS and IRC catalogs simultaneously.

PHP interface programs, {\tt x\_radial.php} 
for Radial Search and {\tt x\_rect.php} for Rectangular Search,
have two operation modes.
One is Single Search Mode, and the other is Multiple Search Mode.

Single Search Mode takes a string parameter
to specify a catalog such as {\tt catalog=fis},
and outputs a result of selected catalog.
Single Search Mode has
an additional argument {\tt contents\_only}.
When the argument {\tt contents\_only=true} is given,
the PHP program does not return the full HTML contents
(beginning {\tt <HTML>}), 
but a part of HTML (HTML of the SQL command and table of results).
This argument is used for Multiple Search Mode
or future services.

When checkboxes FIS and IRC are both selected
(i.e., the URL has {\tt fis=true\&irc=true})
in the Radial Search or Rectangular Search pages,
the PHP program {\tt x\_radial.php} or {\tt x\_rect.php} 
works in Multiple Search Mode.
It forwards the HTTP request to the Single Search Mode of the same
program of the local host with a {\tt contents\_only=true} argument for each
catalog.
The results are returned to the program of Multiple Search Mode,
which attaches the header and footer and outputs into the HTTP stream.
That is, Multiple Search Mode and Single Search Mode
are, in fact, realized by the same program,
and the Multiple Search Mode works like a wrapper program.

Thus, an option to output a part of HTML is useful
to build a tool to search a number of catalogs simultaneously.
It is very easy to build a Web application
that unifies a number of results of radial and rectangular searches
on different servers.%
\footnt{Note that links in HTML contents should be absolute URLs for such a purpose.}
If such a tiny technique is used universally,
we will be able to easily create extensive services without having
many catalog data at a datacenter.

\subsection{Using Images of SkyServer and SkyView in Our CAS}
\label{using_images_in_cas}

Our Explore Tool and Image List Tool provide
quick-look images in multiwavelengths.
Although image products from AKARI All-Sky Survey are not 
available now,
images of SDSS, IRIS, 2MASS, and DSS2
are displayed in them.
These images are not stored in our local server, but in 
the SDSS SkyServer and SkyView.

To inlay the quick-look images into our tools,
the PHP programs {\tt obj.php} and {\tt list.php} 
dynamically generate {\tt <img src= $...$ />} tags
so that users' browsers display the images after downloading them.
SkyServer provides a Web API to obtain a color JPEG image directly
using a single GET method request; therefore, there is no
requirement of additional programs to inlay them.
On the other hand,
SkyView does not provide such a Web API with a single GET method
to get a color image.%
\footnt{There exists an API with a single GET method for
a single-band image.
}
So we created a wrapper program to obtain a color JPEG image 
using a single GET method request.
This is an example of the URL:\\
\verb|   http://darts.isas.jaxa.jp/ir/akari/cas/|\\
\verb|getskyview.php?survey=DSS2+IR%2CDSS2+Red%2|\\
\verb|CDSS2+Blue&rgb=t&position=182.04711396%2C2|\\
\verb|.87884015&scale=0.000277777777777778&pixel|\\
\verb|s=200%2C200&grid=J2000&gridlabels=1|

To create Web-based visual tools easily,
it is very important for servers to provide 
open Web APIs with a single GET method request
to obtain a color image directly in the format supported by the browsers.
If such services appear universally,
we will be able to create attractive visual tools without having
a huge amount of imaging data at each datacenter.
This concept is common for such visual tools and
search tools of multiple catalogs, as described in 
\S \ref{search_multiple_catalogs}.

\subsection{Client-Side Tools for SQL Search on Command-Line}

\begin{figure*}[!t]
\verb|      <?|\\
\verb|        $db = "DR1";|\\
\verb|        $format = "text";|\\
\verb|        $cmd = $argv[1];    /* Obtain SQL statement from command-line args */|\\
\verb|        /* Create URL */|\\
\verb|        $url = "http://darts.isas.jaxa.jp/ir/akari/cas/tools/search/x_sql.php?";|\\
\verb|        $prms = array( 'db'     => $db,|\\
\verb|                       'format' => $format,|\\
\verb|                       'cmd'    => $cmd|\\
\verb|                     );|\\
\verb|        /* Access to AKARI-CAS */|\\
\verb|        $fp = fopen($url . http_build_query($prms), "r");|\\
\verb|        if ( $fp === false ) {|\\
\verb|          fputs(STDERR, "ERROR: fopen() failed\n");|\\
\verb|          exit(1);|\\
\verb|        }|\\
\verb|        /* Read returned catalog data */|\\
\verb|        /* Store the column names */|\\
\verb|        if ( ($s=fgets($fp)) !== false ) {|\\
\verb|          $col_names = explode(",", rtrim($s));|\\
\verb|        }|\\
\verb|        /* Store the values to an associative array and display them */|\\
\verb|        while ( ($s=fgets($fp)) !== false ) {|\\
\verb|          $col_vals = array();|\\
\verb|          $v = explode(",", rtrim($s));|\\
\verb|          for ( $i=0 ; $i < count($v) ; $i++ ) $col_vals[$col_names[$i]] = $v[$i];|\\
\verb|          print_r($col_vals);|\\
\verb|        }|\\
\verb|        fclose($fp);|\\
\verb|      ?>|
\caption{
Example PHP code to access SQL Search of AKARI-CAS.
To use this program, an SQL statement is set to first argument
of the command line for executing this script.}
\label{fig:php_example}
\end{figure*}

Server program {\tt x\_sql.php} for SQL Search has interfaces of both POST and GET.
The POST version is used for input from a browser, and
the GET version is used for command-line tools.
The client-side tools can simply send a single GET method request to
{\tt x\_sql.php} and receive returned results.
Therefore, users can access SQL Search with command-based
HTTP clients such as {\tt wget} or {\tt curl}.
On PHP, using \verb|fopen()| or
\verb|file_get_contents()| with \verb|http_build_query()|
is the most useful way to send their request to `SQL Search'.
See Figure \ref{fig:php_example} for an example PHP code.

AKARI-CAS provides two 
kinds of client-side tools that enable direct access to the database
from users' local computers.
One is sqlcl\_akari.py%
\footnt{Original code (sqlcl.py) was written by T. Budav\'ari,
who gave us permission to modify and re-distribute his program.}
written in Python;
sqlcl\_akari.py can be used by itself from UNIX command shells or from
users' scripts.
Another tool is queryakari.pro written in IDL.
It is 
written as an IDL function that submits an SQL command to AKARI-CAS.
The query results are returned as an array of an IDL anonymous  
structure,
whose tags are the same as the column names of the query results.
This allows users to easily add the querying functionality to the CAS  
to one's own IDL programs.

\section{Summary}
\label{conclusions}

We developed AKARI Catalogue Archive Server (AKARI-CAS) to provide basic
tools to access AKARI-related catalogs for various studies.
Our tools have been developed following the concept of SDSS SkyServer,
and users can search, match up, 
and browse stored data using our attractive tools.

We discussed our direction of implementation for our service
that supports many levels of users.
We concluded that the whole processing of tables should be implemented
into an RDBMS layer.
We found that PostgreSQL is the best open-source RDBMS product for our
requirements, and we developed various facilities using stored functions of
PL/pgSQL, C, and SQL in the database.
In contrast, the Web application layer became compact;
this minimizes the cost of maintenance in the long term.

We presented our simple techniques 
without HTM or HEALPix
to perform fast radial search and dynamic cross-identification
using RDBMS.
With our report of the security configuration of PostgreSQL, 
the information
in this article will be helpful to develop future astronomical
Web-based database systems.
In the development of the Web application, we showed our idea to create
extensive services using only standard technology and without having
a huge amount of data at a datacenter.
Some thought put into the definition of Web APIs and public release of the
APIs can increase the possibility of online astronomical services.

\acknowledgments

We are grateful to Aniruddha R. Thakar and Yanxia Zhang
for informing us about status of their cross-identification and
literature.
We thank anonymous referee for useful comments and suggestions.

This research has made use of the SIMBAD database, operated at CDS, 
Strasbourg, France. 
This research has made use of the NASA/IPAC Extragalactic Database (NED) 
which is operated by the Jet Propulsion Laboratory, California Institute of 
Technology, under contract with the National Aeronautics and Space 
Administration. 
This research has made use of the NASA/IPAC Infrared Science Archive (IRSA) 
which is operated by the Jet Propulsion Laboratory, California Institute of 
Technology, under contract with the National Aeronautics and Space 
Administration.
We acknowledge the use of NASA's SkyView facility 
located at NASA Goddard Space Flight Center.

Funding for the SDSS and SDSS-II has been provided by the Alfred P. Sloan Foundation, the Participating Institutions, the National Science Foundation, the U.S. Department of Energy, the National Aeronautics and Space Administration, the Japanese Monbukagakusho, the Max Planck Society, and the Higher Education Funding Council for England. The SDSS Web Site is http://www.sdss.org/.
The SDSS is managed by the Astrophysical Research Consortium for the Participating Institutions. The Participating Institutions are the American Museum of Natural History, Astrophysical Institute Potsdam, University of Basel, University of Cambridge, Case Western Reserve University, University of Chicago, Drexel University, Fermilab, the Institute for Advanced Study, the Japan Participation Group, Johns Hopkins University, the Joint Institute for Nuclear Astrophysics, the Kavli Institute for Particle Astrophysics and Cosmology, the Korean Scientist Group, the Chinese Academy of Sciences (LAMOST), Los Alamos National Laboratory, the Max-Planck-Institute for Astronomy (MPIA), the Max-Planck-Institute for Astrophysics (MPA), New Mexico State University, Ohio State University, University of Pittsburgh, University of Portsmouth, Princeton University, the United States Naval Observatory, and the University of Washington.

S.T. is financially supported by the Japan Society for the Promotion 
of Science.

\section*{Appendix A: Basic concepts and notes of the AKARI-CAS Web API}
 

AKARI-CAS employs stateless design and defines Web API.
An API returns a search result in HTML or text format.
Input parameters are basically taken with the GET method of
HTTP protocol;
however, the POST method is also applied, 
as required by some tools.

To search catalogs or images,
users set their search conditions on the arguments of APIs.
In typical use with Radial Search or Explore,
after setting the {\tt catalog} parameter to choose a catalog,
positional values such as longitude and latitude
can be set on {\tt lon} and {\tt lat} parameters
in degrees or sexagesimal.
The coordinate system of {\tt lon} and {\tt lat} parameters
can be selected by the {\tt coordinate} parameter;
{\tt j2000}, {\tt b1950}, {\tt ecliptic}, and {\tt galactic} are
supported.
Instead of position,
users can input an object name that can be resolved by SIMBAD or NED
with the {\tt ident\_simbad} or {\tt ident\_ned} parameters, respectively.
In some parameters,
Boolean values should be set in {\tt true}/{\tt false} or
{\tt 1}/{\tt 0}.

All 
APIs use the {\tt db} parameter
to select the database that users want to use.
A database only has a version of a catalog set;
for example, the DR1 database has FIS version 1.0 and IRC version 1.0
catalogs.

APIs for the table output support 
display of the column types (e.g., int4, bpchar, etc.) in the table header.
To enable this feature, use the {\tt types=true} argument.
This is useful to create users' programs in certain languages.
The output format of a table is HTML or text
in these tools.
Users can choose delimiter and newline characters
in the text format with {\tt delimiter} and {\tt newline}
parameters.


We will keep our Web API simple.
To perform a search with a complicated condition,
users can learn SQL with our SQL Tutorial page and
use the API of SQL Search.

See the Web API page in the Help category of our AKARI-CAS
for more information.

\section*{Appendix B: Security Configuration of PostgreSQL}

We show step-by-step procedures in the script described in
\S \ref{pgsql_security}
for security configuration of PostgreSQL in the AKARI-CAS:
\renewcommand{\labelenumi}{\arabic{enumi}.}
\begin{enumerate}
 \item
      Revoke permission of all schemata from public user:

      {\small \verb|REVOKE ALL ON SCHEMA public from public;|}\\
      {\small \verb|REVOKE ALL ON SCHEMA pg_catalog from public;|}

 \item
      Grant permission of all schemata from admin user:

      {\small \verb|GRANT ALL ON SCHEMA public to admin;|}\\
      {\small \verb|GRANT ALL ON SCHEMA pg_catalog to admin;|}
 \item
      Create guest schema:

      {\small \verb|CREATE SCHEMA guest;|}

 \item
      Grant read-only permission of all schemata to guest user:

      {\small \verb|REVOKE ALL ON SCHEMA guest from guest;|}\\
      {\small \verb|REVOKE ALL ON SCHEMA public from guest;|}\\
      {\small \verb|REVOKE ALL ON SCHEMA pg_catalog from guest;|}\\
      {\small \verb|GRANT USAGE ON SCHEMA guest to guest;|}\\
      {\small \verb|GRANT USAGE ON SCHEMA public to guest;|}\\
      {\small \verb|GRANT USAGE ON SCHEMA pg_catalog to guest;|}

 \item
      Revoke permission of all relations in pg\_catalog schema from
      public user:

      {\small {\tt REVOKE ALL ON pg\_}{\it all } {\tt FROM public;}}

 \item
      Grant permission of all relations in pg\_catalog schema to
      admin user:

      {\small {\tt GRANT SELECT ON pg\_}{\it all } {\tt TO admin;}}

 \item
      Grant permission of pg\_type to guest user:

      {\small {\tt GRANT SELECT ON pg\_type TO guest;}}

      This is required to access type information of returned
      result of an SQL execution from Web applications.
 \item
      Revoke permission of all functions in pg\_catalog schema from
      public user:

      {\small {\tt REVOKE ALL ON} {\it all}{\tt ( $...$ ) FROM public;}}
 \item
      Grant permission of all functions in pg\_catalog schema to
      admin user:

      {\small {\tt GRANT ALL ON} {\it all}{\tt ( $...$ ) TO admin;}}

 \item
      Grant permission of selected functions in pg\_catalog schema to
      guest user:

      {\small {\tt GRANT EXECUTE ON} {\it selected}{\tt ( $...$ ) TO guest;}}

      We have to select functions carefully.
\end{enumerate}



\label{lastpage}

\end{document}